\documentclass{PoS}

\title{Multi-Lepton Events at HERA}

\ShortTitle{Multi-Lepton Events at HERA}

\author{\speaker{David M. South}\thanks{on behalf of the H1 and ZEUS Collaborations}\\
        Deutsches Elektronen Synchrotron,\\
        Notkestrasse 85, 22607 Hamburg, Germany\\
        E-mail: \email{david.south@desy.de}}

\abstract{The analysis of events containing multiple high $P_{T}$
leptons (electrons and muons) produced in $e^{\pm}p$ collisions
has been performed with the H1 and ZEUS detectors at HERA,
using the full data sets collected by the experiments
in the period $1994-2007$.
Mutually exclusive event topologies containing at least two
charged leptons are analysed.
The H1 and ZEUS data, corresponding to a total integrated luminosity
of about $1$~fb$^{-1}$, are combined in a common phase space.
The observed event yields are compared to the predictions from
the Standard Model.
In general a good agreement is found, where the expectation is
dominated by photon--photon collisions.
Interesting events at high mass and high $P_{T}$ are observed by
both experiments.
The total and differential cross sections for multi--lepton
production at HERA are also measured.}

\FullConference{35th International Conference of High Energy Physics - ICHEP2010,\\
		July 22-28, 2010\\
		Paris France}

\begin{document}

\section{Introduction}

At HERA, protons with an energy up to $920$~GeV were brought into
collision with electrons or positrons of energy $27.6$~GeV at two
experiments, H1 and ZEUS, each of which collected about
$0.5$~fb$^{-1}$ of data in the period $1994$--$2007$.
The collisions produced at HERA at a centre of mass energy of
up to $319$~GeV provide an ideal environment to study rare processes,
set constraints on the Standard Model (SM) and search for new
particles and physics beyond the Standard Model (BSM).
The good lepton identification and hadronic final state reconstruction
of the H1 and ZEUS experiments means that such topologies provide a
clean signal, and the high mass, high $P_{T}$ regions, where the SM
expectation is low, may be investigated for signs of new physics.
Measurements of both multi--electron~\cite{h1multielectron} and
muon pair~\cite{h1multimuon} production at high transverse momentum
were performed by the H1 Collaboration using their HERA I data sample
($1994$--$2000$), which corresponds to an integrated luminosity of
$115$~pb$^{-1}$.
Both H1~\cite{h1multilep} and ZEUS~\cite{zeusmultilep} have now
published analyses of multi--lepton events using their complete data sets.
A combined analyses of the complete HERA data has also been
performed~\cite{h1zeusmultilep}.

\section{Multi-lepton Events at HERA}

The production of multi--lepton final states in $e^{\pm}p$ collisions
proceeds in the SM mainly via photon--photon interactions.
Multi--lepton events are simulated using the
GRAPE~\cite{Abe:2000cv} Monte Carlo (MC), which includes all
electroweak matrix elements at tree level.
The amount of SM background present in the analysis depends on the
number and flavour of the identified leptons in the event.
Neutral current deep inelastic scattering (NC DIS, $ep\rightarrow eX$) and
QED Compton scattering (QEDC, $ep \to e\gamma X$)
events contribute as background to multi--lepton events with at least one
identified electron in the final state.
The background contribution to the number of events with two identified
muons or more than two identified leptons is negligible.


Electrons are identified in the polar-angle range
$5^\circ < \theta_{e} < 175^\circ$ with $E_{e}>10$~GeV in
the range $5^\circ < \theta_{e} < 150^\circ$ and $E_{e}>5$~GeV
in the backward region ($150^\circ < \theta_{e} < 175^\circ$).
In the H1 analysis~\cite{h1multilep}, electrons
with an energy $E_{e}>5$~GeV are also allowed in the
forward region ($5^\circ < \theta_{e} < 20^\circ$).
Muon candidates are identified in the range
$20^\circ <\theta_{\mu} < 160^\circ$ with $P_{T}^{\mu}>2$~GeV.
All lepton candidates are required to be isolated with respect to each
other, as well as other calorimeter deposits and tracks in the event,
by a minimum distance of at least $0.5$ units in the $\eta-\phi$ plane.

At least two central ($20^\circ < \theta < 150^\circ$) lepton candidates
are required, one of which must have $P_T^\ell>10$~GeV and
the other $P_T^{\ell}>5$~GeV. 
Additional leptons identified in the detector according to the
criteria defined above may be present in the event.
According to the number and the flavour of the lepton candidates,
the events are classified into mutually exclusive topologies, such as
$ee$, $eee$, $e\mu$, $e\mu\mu$, $\mu\mu$ and so on.
A full description of the event selection is presented in~\cite{h1zeusmultilep}.


\section{Results}

\begin{figure*}[h]
\centering
\includegraphics[width=0.445\columnwidth]{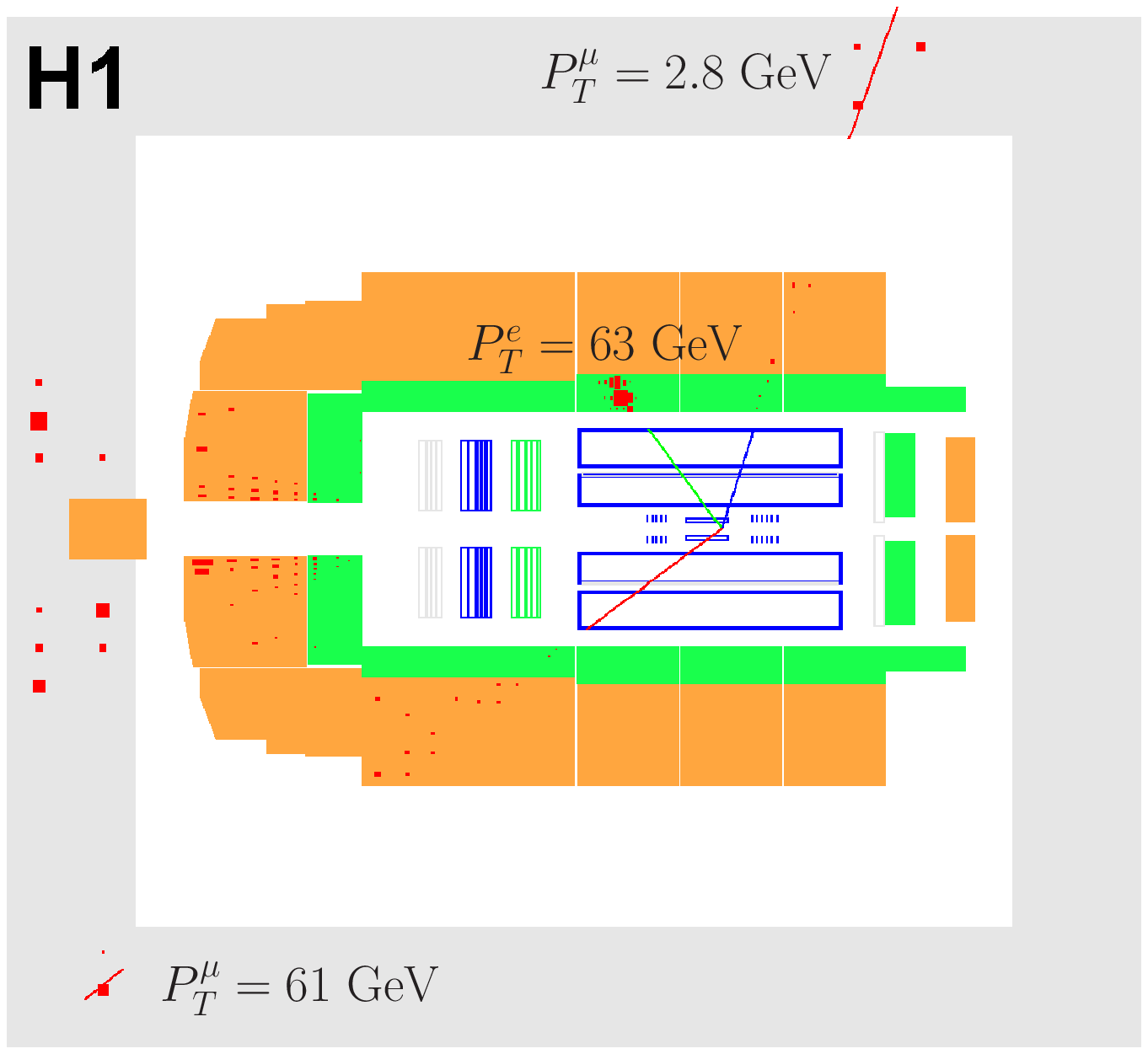}
\includegraphics[width=0.435\columnwidth]{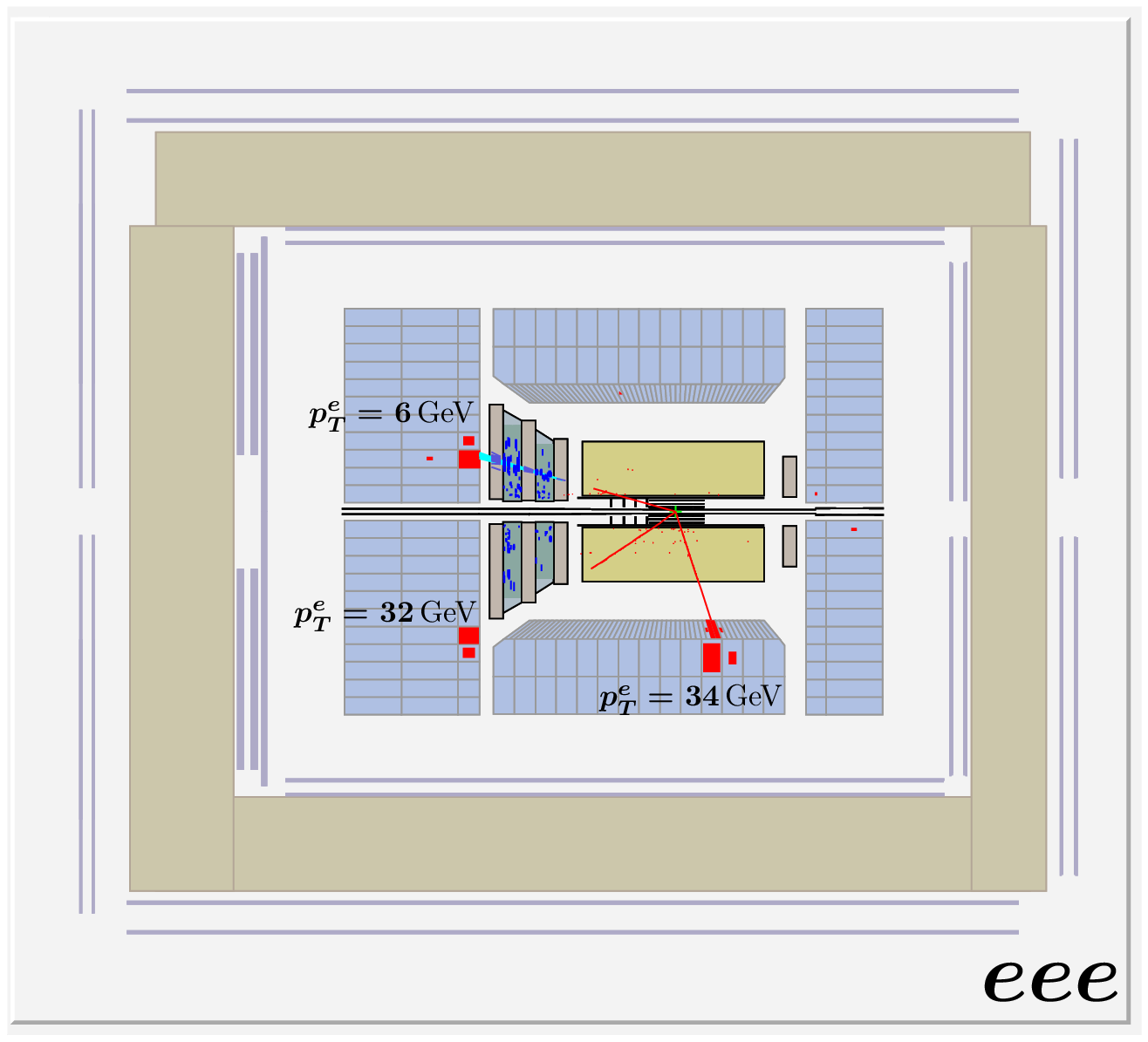}
\vspace{-0.35cm}
\caption{High mass events observed in the H1 (left) and ZEUS (right)
multi--lepton analyses. In the H1 $e\mu\mu$ event, the invariant mass
$M_{12}$ of the electron and the highest $P_{T}$ muon is $127$ GeV. In the ZEUS
$eee$ event, the invariant mass $M_{12}$ of the two highest $P_{T}$
electrons is $113$ GeV.}
\label{fig:events}
\end{figure*}

In the H1 and ZEUS analyses the number of observed events in the various
event topologies is found in general to be in good agreement with the SM
prediction~\cite{h1multilep,zeusmultilep}.
The SM prediction is dominated by multi--lepton events, and only the
$ee$ toplogy contains significant SM background, arising from NC DIS
and QEDC processes.
Interesting events are however observed by both H1 and ZEUS at
high invariant mass $M_{12}$, which is constructed from the highest
two $P_{T}$ leptons in the event.
Two such events are displayed in figure~\ref{fig:events}.
Furthermore, in the H1 analysis five events are observed in the
kinematic region $\sum{P_T}>100$~GeV compared to a SM
expectation of $1.60\pm 0.20$.
All five events are observed in the $e^{+}p$ data, compared to a SM
expectation of $0.96\pm 0.12$.

\begin{figure*}[h]
\centering
\includegraphics[width=0.44\columnwidth]{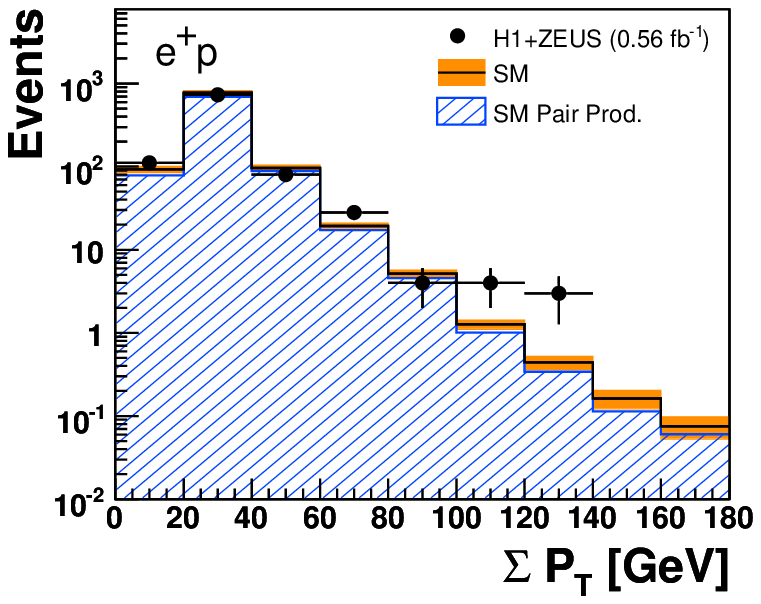}
\includegraphics[width=0.44\columnwidth]{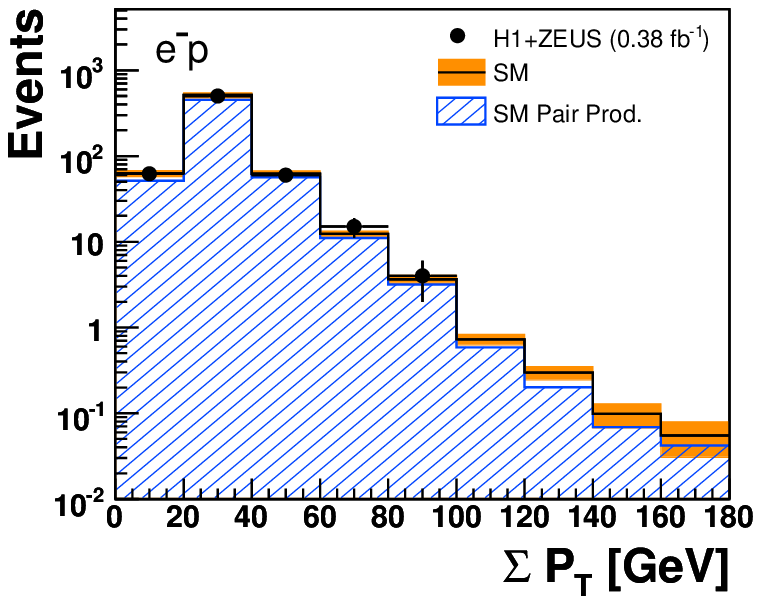}
\vspace{-0.35cm}
\caption{The distribution of the scalar sum of the transverse momenta
$\Sigma P_{T}$ for the combined di--lepton and tri--lepton events in
the complete HERA $e^{+}p$ (left) and $e^{-}p$ (right) data. The points 
are the data and the full histogram is the total SM expectation, where the
shaded band indicates the uncertainty on the SM prediction. The signal
component of the SM expectation is shown by the striped histogram.}
\label{fig:mleppts}
\end{figure*}

An analysis of the full HERA data sample, corresponding to an integrated
luminosity of $0.94$~fb$^{-1}$, is performed in a common phase space.
The electron energy threshold is increased in the H1 analysis
from $5$ to $10$~GeV in the forward region.
Both the number of the observed events and the cross sections for multi--lepton
production measured by the two experiments are combined, allowing an increased
sensisivity to rare processes in the high mass and high $P_{T}$ regions and an
improved precision in the measured cross section.
Once again, in general a good overall agreement is found between the data and the
SM predictions.
However, for $\sum{P_T}>100~{\rm GeV}$, seven events are observed in the data,
compared to $3.13\pm0.26$ expected from the SM.
In addition to the five events from H1 described above, two further data events are
observed in this region by ZEUS.
The combined $\sum{P_T}$ distributions for the full HERA data are
shown in figure~\ref{fig:mleppts}, separately for the $e^{+}p$ and $e^{-}p$ data.
It can seen that all of the high $\sum{P_T}$ events are recorded in the
$e^{+}p$ data, where seven events are observed compared to a SM
expectation of $1.94\pm 0.17$.

\begin{figure*}[h]
\centering
\includegraphics[width=0.88\columnwidth]{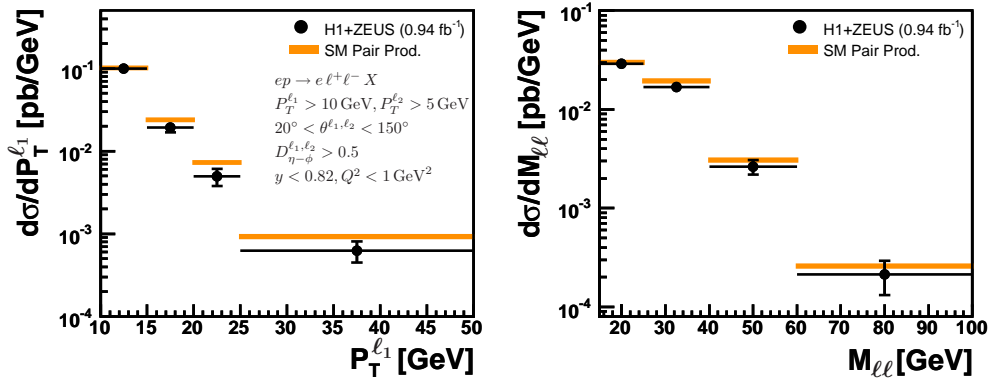}
\vspace{-0.35cm}
\caption{The cross section for lepton--pair photoproduction
as a function of the leading lepton transverse momentum $P^{{\ell}_{1}}_{T}$ (left)
and the invariant mass of the lepton pair $M_{\ell\ell}$ (right).
The error on the data (points) represents the statistical and systematic
uncertainties added in quadrature. The shaded band represents the uncertainty
on the SM prediction.}
\label{fig:mlepxs}
\end{figure*}

The lepton--pair production cross section is measured in a
phase space dominated by photon--photon interactions as
$0.66 \pm 0.03 ({\rm stat.}) \pm 0.03 (\rm sys.)$~pb,
in agreement with the SM prediction of $0.69 \pm 0.02$~pb.
The cross section is also measured as a function of the transverse
momentum of the leading lepton, $P^{{\ell}_{1}}_{T}$ and the invariant
mass of the lepton pair $M_{\ell\ell}$, as shown in figure~\ref{fig:mlepxs}.

\section{Conclusions}

The final analyses of multi--lepton events at HERA have been completed,
including a combination of the full H1 and ZEUS data.
In general a good agreement is observed between the data and the SM
expectation.
Events are observed by both H1 and ZEUS with high invariant mass of the two
highest transverse momentum leptons, $M_{12}>100~{\rm GeV}$, and a high
scalar sum of the lepton transverse momenta, $\sum{P_T}>100~{\rm GeV}$,
but only in the $e^+p$ collision data.
The total and differential cross sections for multi--lepton production at
HERA are also measured.

\end{document}